# Microwave Surface Resistance and Upper-Critical-Field Anisotropy of MgB$_2$ Superconductor

A. Agliolo Gallitto, S. Fricano and M. Li Vigni

*INFM and Dipartimento di Scienze Fisiche e Astronomiche, University of Palermo, via Archirafi 36, 90123, Palermo, Italy*

**Abstract**

The field-induced variations of the microwave surface resistance are investigated in two different samples of powdered MgB$_2$. The experimental results can be justified in the framework of the Coffey and Clem model with fluxons moving in the flux-flow regime, provided that the anisotropy of the upper critical field is taken into due account. Assuming the angular dependence of the upper critical field expected from the anisotropic Ginzburg-Landau theory, we determine the anisotropy factor by fitting the experimental data of the field dependence of the microwave surface resistance. We show that the anisotropy factor is constant in a range of temperatures of about 3 K below $T_c$ and it takes on different values for the two samples, even though they have similar shape.

**Corresponding Author:** A. Agliolo Gallitto

Dipartimento di Scienze Fisiche e Astronomiche – Via Archirafi 36, 90123 Palermo, Italy

Tel: +39 091 6234 207 – Fax +39 091 6162461 – E-mail: agliolo@fisica.unipa.it

**Introduction**

Since the discovery of the superconductivity at 40 K in magnesium diboride [1], a large number of studies have been devoted to determine specific properties of this compound as well as to understand the origin of the superconducting state. In the literature, there is not general consensus about both the interpretation of experimental results and the values of characteristic parameters of the superconducting state, such as penetration depth, normal-state resistivity, critical fields [2]. This suggests that the measured parameters depend on the sample quality and growth method. Several properties, such as the anisotropy of the upper critical field and the upward curvature of $H_{c2}(T)$, have been ascertained [3-16]. However, the anisotropy coefficient $g$ is not yet established; the reported values spread over a wide range, from 1.1 to 13. The values of both $g$ and $H_{c2}$ reported in the literature depend on the measurement method [5, 7, 15], the criterion for defining the upper critical field [13, 16] and the kind of sample, i.e., single crystal [3-7], film [11-13], aligned crystallites [14] and powder [9, 10]. It has been highlighted a temperature dependence of $g$ [6, 7, 9, 11], which cannot be justified in the framework of the anisotropic Ginzburg-Landau (AGL) theory. Nevertheless, by using the angular dependence of $H_{c2}$ expected from the AGL theory several experimental results have been accounted for quite well in a large range of temperatures, applied fields and field orientations [6, 10, 11].

Measurements of the microwave surface resistance, $R_S$, allow investigating the mechanisms responsible for microwave energy losses in superconductors as well as determining specific characteristics of the samples. Furthermore, investigation of the magnetic field dependence of $R_S$ may provide important information on the fluxon dynamics in the mixed state [17-20]. In the last year, several papers have reported on the temperature dependence of the surface impedance of $MgB_2$, but only a few of them have concerned the magnetic field dependence [21-23]. It has been shown that the field-induced variations of the microwave surface resistance of $MgB_2$ are more enhanced than those observed in conventional as well as in cuprate superconductors. This agrees with several studies reported in the literature, which show that the presence of the magnetic field strongly affects thermal, electric and magnetic properties of the $MgB_2$ superconductor [3, 5, 15, 16, 24].

In this paper we investigate the magnetic-field-induced variations of the microwave surface resistance in two different samples of $MgB_2$: one made up of Alfa-Aesar powder and the other obtained by milling a bulk ceramic synthesized by direct reaction from elements [25]. The microwave surface resistance has been measured using the cavity perturbation technique at 9.6 GHz. It has been investigated in the range of temperatures 4.2 K ÷ $T_c$, at different values of the static magnetic field up to 10 kOe. The results show that the microwave losses are strongly affected by the magnetic field in all the range of temperatures investigated, even for relatively low field values. Our attention will be mainly devoted to the investigation at temperatures close to $T_c$. The experimental results are discussed in the framework of the Coffey and Clem model [18]. We show that, near $T_c$, the field-induced variations of $R_S$ can be justified by taking into account the losses due to the presence of the normal fluid and the motion of fluxons in the flux-flow regime, provided that the anisotropy of the upper critical is properly considered. The anisotropy factor ***g*** has been determined by fitting the experimental data under the hypothesis that the angular dependence of $H_{c2}$ follows the law expected from the AGL model. ***g*** results to be constant in a range of temperatures of ~ 3 K below $T_c$ and it takes on different values for the two samples, even though they have similar shape.

**Experimental results and samples**

The microwave surface resistance has been investigated in two different $MgB_2$ samples. One (P#1) consists of 10 mg of Alfa-Aesar powder. The other (P#2) has been obtained by milling a bulk piece of ceramic $MgB_2$, which had been synthesized by the direct reaction of boron powder with a lump of magnesium metal, both of purity better than 99.95% [25]. The powdering process of the P#2 sample has been performed in order to increase its effective surface, improving the measurement sensitivity, and make clearer the conditions for a comparison of the results in samples of similar shape. The two samples have equal weight and are both sealed in Plexiglas holders. From measurements performed in a bulk specimen of the same batch of P#2, the normal-state resistivity and the residual surface resistance at $T = 0$ have been determined: $r(T_c) \approx 15$ μΩcm and $R_{res}(0) \approx 6$ mΩ [26].

The experiments have been performed using the cavity perturbation technique. The cavity, of cylindrical shape with golden-plated walls, was tuned in the $TE_{011}$ mode resonating at 9.6 GHz. Its $Q$-factor is ~ 25000 at room temperature and ~ 40000 at the helium temperature. The sample is located at the bottom of the cavity, in a region where the microwave magnetic field is maximal, and placed between the expansions of an electromagnet, which generates fields up to ~ 10 kOe. Two additional coils, externally fed, allow reducing to zero the residual field of the electromagnet and working at low magnetic fields. The $Q$-factor of the cavity has been measured by means of an *hp*-8719D Network Analyser. The surface resistance, $R_S$, of the sample is proportional to $(1/Q_L - 1/Q_U)$, where $Q_U$ is the $Q$-factor of the empty cavity and $Q_L$ is that of the loaded cavity. In order to disregard the geometric factor of the sample, it is convenient to normalise the deduced values of $R_S$ to the value of the surface resistance at a fixed temperature in the normal state, $R_N$. Measurements have been performed as a function of the temperature and the static field, $H_0$. All the results have been obtained with the static magnetic field normal to the microwave current.

Fig. 1 shows $(1/Q_L - 1/Q_U)$ as a function of the temperature in the absence of external magnetic field, for both samples. Since the samples have roughly the same shape, we can say that the normal-state surface resistance of P#1 is about five times greater than that of P#2. Furthermore, the low-$T$ residual surface resistances of the two samples differ from each other by a factor two.

In Fig. 2 we report the temperature dependence of $R_S/R_N$ of P#1, at different values of the static field. The results have been obtained according to the following procedure: the sample was zero-field cooled down to 4.2 K, then $H_0$ was set at a given value, which was kept constant during the time in which all measurements in the temperature range 4.2 ÷ 40 K have been performed. $R_S$ has been normalized to its value at $H_0 = 0$ and $T = 40$ K. Similarly to what occurs in high-$T_c$ superconductors, on increasing $H_0$ the $R_S(T)$ curve broadens and shifts toward lower temperatures. However, the effect of the magnetic field in the $R_S(T)$ curves of $MgB_2$ is more significant than that reported in the literature for cuprate superconductors [19, 20, 27]. In particular, an unusually enhanced field dependence of the low-$T$ residual surface resistance has been detected.

By measuring the shift of the transition temperature induced by $H_0$ we have deduced the temperature dependence of the upper critical field near $T_c$. The deduced values of $H_{c2}(T)$ are shown in Fig. 3, for both P#1 (squares) and P#2 (triangles) samples. The inset shows the procedure followed to determine $H_{c2}(T)$. Both the $H_{c2}(T)$ curves show values of $(dH_{c2}/dT)_{T=T_c}$ of about 2 kOe/K and exhibit an upward curvature, in agreement with what reported in the literature [2, 8, 9, 15, 16].

Figs. 4 (a) and 4 (b) show the normalized values of the surface resistance as a function of the static magnetic field for P#1 and P#2, respectively, at different values of the temperature, near $T_c$. As expected, on increasing the temperature the samples go into the normal state at lower values of the magnetic field. A comparison between panels (a) and (b) shows that the values of $R_S(T)/R_N$ of P#1 are smaller than those of P#2. Furthermore, although the two samples have approximately the same values of $H_{c2}(T)$, on increasing $H_0$, the surface resistance of P#2 approaches the normal-state value faster than that of P#1.

**Discussion**

Microwave losses induced by static magnetic fields in high-$T_c$ superconductors have been investigated by several authors [18-20, 30]; the losses have been mainly ascribed to the motion of fluxons in the flux-flow regime. Our results show that the effect of the applied magnetic field on the microwave surface resistance of $MgB_2$ is stronger than that observed in both conventional and cuprate superconductors, in all the range of temperatures $T < T_c$. It has already been shown that, in $MgB_2$ superconductors, the field-induced variations of $R_S$ at low temperatures follow a $\sqrt{B}$ law [22, 23], as expected in the framework of Coffey and Clem model in the pinning limit [17, 19]. We have suggested [23] that the enhanced variation can be ascribed to several properties of the $MgB_2$ compound, such as the weak-pinning effect [15, 28, 29] and the relatively small values of both normal-state resistivity [2, 15, 26] and upper critical field [2].

In this paper, our aim is to discuss the results obtained at temperatures close to $T_c$. In this range, the field-induced variations of the microwave surface impedance have mainly been ascribed to the motion of fluxons induced by the microwave current. However, it has been pointed out that a noticeable contribution can arise by the presence of the normal fluid [18, 30], especially when the external magnetic field is of the same order of magnitude of the upper critical field and/or at temperatures close to $T_c$.

Coffey and Clem [18] have developed a comprehensive theory for the electromagnetic response of type-II superconductors in the mixed state, taking into account fluxon motion and pinning effects in the two-fluid model of the superconductivity. The theory applies for $H_0 > 2H_{c1}$, when the static field inside the sample can be supposed as generated by an uniform density of fluxons; in this case $H_0 \approx B_0 = n_0 \boldsymbol{f}_0$. The authors show that the electromagnetic field inside the sample is characterized by a complex penetration depth, $\tilde{\lambda}$, which is affected by the fluxon motion as well as by the very presence of vortices, which bring along normal material in their cores. In the linear approximation, $H(\omega) \ll H_0$, they found the following expression for $\tilde{\lambda}$:

$$\tilde{\lambda}^2 = \frac{\lambda^2 + i\delta_v^2/2}{1 - 2i\lambda^2/\delta^2}, \tag{1}$$

where $\delta_v$ is the complex effective skin depth arising from vortex motion; $\lambda$ and $\delta$, the London penetration depth and the normal-fluid skin depth, are given by

$$\lambda = \frac{\lambda_0}{\sqrt{(1-w)(1-B_0/B_{C2})}} \qquad \delta = \frac{\delta_0}{\sqrt{1-(1-w)(1-B_0/B_{C2})}}; \tag{2}$$

here, $\lambda_0$ is the London penetration depth at $T = 0$; $\delta_0$ is the normal skin depth at $T = T_c$; $w$ is the fractions of normal electrons at $H_0 = 0$, in the framework of the two-fluid model $w = (T/T_c)^4$.

Because of the weak-pinning effect, the irreversibility line in MgB$_2$ lies well below $H_{c2}(T)$ [15, 28, 29]; as a consequence of this, fluxons are in the viscous regime in a wide region of the $H$-$T$ phase diagram. Therefore, in the ranges of fields and temperatures at which we are interested, it is reasonable assuming that vortices move in the flux-flow regime, especially at so high frequency. In this case, $\delta_v$ is given by

$$\delta_v = \frac{B_0 \phi_0}{2\pi \eta \omega}, \tag{3}$$

where $\eta$ is the viscous drag coefficient, $\eta = \phi_0 B_{c2}(T)/c^2 \rho_n$.

In the London local limit, the surface impedance is given by

$$Z_S = R_S - iX_S = i\omega\tilde{\lambda}; \tag{4}$$

thus, in the Coffey and Clem model, the field dependence of surface impedance arises from changes of the complex penetration depth, induced by the presence of fluxons and their motion. The parameters necessary to perform a comparison between the experimental and expected results are the $H_{c2}(T)$ values, which we have deduced from the experimental data, and the ratio $l_0/d_0$. It is easy to see that $l_0/d_0 = \sqrt{(\omega t/2)}$, where $t$ is the scattering time of the normal electrons. Values of $t$ reported in the literature for ceramic $MgB_2$ are $\sim 10^{-13}$ s [26]; so, at microwave frequencies and for these samples the ratio $l_0/d_0$ could be of the order of $10^{-2}$. The dashed line of Fig. 4 (b) shows the expected curve obtained using Eqs. (1) ÷ (4) with $l_0/d_0 = 10^{-2}$ and $H_{c2} = 6$ kOe. As one can see, for $H_0 \ll H_{c2}$, the experimental curve $R_S(H_0)$ varies faster than the expected one; on the contrary, for $H_0 \sim H_{c2}$, the experimental curve varies more slowly than the theoretical one. We have calculated curves of $R_S(H_0)$ using for $l_0/d_0$ values ranging from $10^{-1}$ to $10^{-3}$ and for $H_{c2}(T)$ the values deduced from the experimental data, letting them vary within the experimental accuracy. The results have shown that the model, in the present form, cannot account for the experimental data. However, the results can be justified quite well by taking into due account the upper-critical-field anisotropy.

The criterion used for deducing the $H_{c2}(T)$ values reported in Fig. 3 allows to determine the field values at which the whole sample goes into the normal state, i.e., the values of $H_{c2}$ of the crystallites having the c-axis perpendicular to the applied magnetic field. From these considerations, one can qualitatively justify the data of Fig. 4. Indeed, because our samples consist of randomly orientated grains, on increasing the applied field, before it reaches the $H_{c2}^{\perp c}$ value, parts of the sample go through the normal state. For this reason, in the superconducting state the experimental values of $R_S(H_0)$ are larger than the expected ones.

For superconducting materials in which the coherence length is much larger than the periodicity of the crystal lattice, such as the $MgB_2$ compound, one can reasonably assume that the angular dependence of the critical field follows the AGL theory:

$$H_{c2}(\theta) = H_{c2}^{\perp c} / [\gamma^2 cos(\theta)^2 + sin(\theta)^2]^{1/2}, \tag{5}$$

where $\gamma = H_{c2}^{\perp c}/H_{c2}^{\parallel c}$ is the anisotropy factor and $\theta$ is the angle between the c-axis of crystallites and the external magnetic field.

The angular dependence of $H_{c2}$ in MgB$_2$ has been reported in the literature for single crystals [4, 6] and films [11]. The results obtained by Angst et *al.* [6] in single crystals show that the $H_{c2}(\theta)$ values, deduced from torque magnetometry, follow quite well the AGL $H_{c2}(\theta)$ law for all the orientations. On the contrary, data reported in Ref. [4], for single crystals, and in Ref. [11], for films, shows deviations from Eq. (5) at small angles. The AGL theory cannot surely justify the temperature dependence of the anisotropy factor $\gamma$ observed in MgB$_2$ by several authors [6, 9, 11]. Nevertheless, it has been shown that data obtained by different techniques can be well justified using Eq. (5) in a large range of temperatures, magnetic fields and angles [6, 10, 11]. In order to quantitatively account for our experimental data we have assumed $H_{c2}(\theta)$ of Eq. (5). On the other hand, in randomly oriented crystallites possible deviations from Eq. (5) at small angles cannot strongly affect the results.

The expected $R_S(H_0)$ curves depend on $H_{c2}$, as given in Eqs. (1) ÷ (4). Therefore, in order to take into account the effect of the $H_{c2}$ anisotropy in the field dependence of the surface resistance, we have to average the expected $R_S(H_0, H_{c2}(\theta))$ curves over a suitable distribution of the crystallite orientations. In view of the fact that our samples are made up of fine powder, we have assumed a random orientation of the crystallite c-axes with respect to the applied magnetic field. On this hypothesis, the distribution function of the grain orientation will be

$$dN(\theta) = N_0 \sin(\theta) d\theta/2 , \qquad (6)$$

where $N_0$ is the total number of crystallites.

The expected $R_S(H_0)$ curves depend on the values of $H_{c2}^{\perp c}(T)$, $l_0/d_0$ and $\gamma$. In order to fit the data, we have used for $H_{c2}^{\perp c}(T)$ the values deduced from the experimental results, letting them vary within the experimental uncertainty, and we have taken $l_0/d_0$ and $g$ as parameters. However, we have found that the expected results are little sensitive to variations of $l_0/d_0$ as long as it takes on values smaller than 0.1. Since higher values of this parameter are not reasonable, $l_0/d_0$ cannot be determined by fitting the data; so, the only free parameter is $g$

In order to obtain the expected $R_S(H_0)$ curves, we have used $l_0/d_0 = 10^{-2}$. Fittings of the experimental data performed on varying the $g$ value have shown that, in the range of temperatures investigated, $g$ does not depend on temperature, for both samples. The best-fit curves have been obtained using different values of $g$ for the two samples. In particular, we have obtained $g = 3 \pm 0.1$ for P#1 and $g = 5 \pm 0.3$ for P#2. The continuous lines of Fig. 4 are the best-fit curves of the experimental data.

At present, two problems concerning the upper-critical-field anisotropy of MgB$_2$ are open. One concerns the microscopic origin of the temperature dependence of $g$; indeed, it has widely been shown that $g$ decreases on increasing the temperature [6, 9, 11]. This behaviour is related to the fact that $H_{c2}^{\perp c}(T)$ shows an upward curvature not present in $H_{c2}^{\parallel c}(T)$ [4-7, 11-13]. The $g(T)$ feature has been tentatively ascribed to the peculiar double-gap structure of MgB$_2$, recently well characterized [31], but the question is still open [6, 7, 11]. A temperature dependence of the upper-critical-field anisotropy has been already observed in borocarbides [32], although it is more enhanced in MgB$_2$. As shown, in clean superconductors it can be justified using non-local extension of the Ginzburg-Landau equations [32]. Our results show that the investigated samples do not exhibit detectable variations of the upper-critical-field anisotropy in a range of about 3 K below $T_c$. This finding is probably due to the fact that non-local effects are not important at temperatures close to $T_c$.

Another issue largely discussed is the reason of the wide spreading of the $g$ value reported in the literature. The reported values of the anisotropy factor range from 1.1 to 13, depending on the measurement method and kind of sample. The values of *g* determined by resistivity measurements are usually smaller than those deduced by thermal and magnetic measurements [5, 7]. In any case, the value of the anisotropy factor seems to be strongly dependent on the kind of sample [3-14]. In randomly oriented powder, it has been found *g* = 6 – 9 at low temperature and *g* ≈ 3 close to $T_c$ [9, 10]. The *g* value we have obtained for P#1 agrees with those reported by several authors in powdered [9, 10] as well as in polycrystalline [8] samples, on the contrary for P#2 we have obtained a higher value. So, although the two investigated samples have similar shape and they have been investigated by the same measurement method, we have found different values of *g*. This finding corroborates the idea that the anisotropy of the upper critical field depends on the sample growth method, probably because of the different contaminating impurities.

In conclusion, we have investigated the field-induced variations of the microwave surface resistance in two powdered samples of $MgB_2$, grown with different techniques. The field dependence of $R_S$ is enhanced when compared to that detected in both conventional and high-$T_c$ superconductors. We have shown that at temperatures close to $T_c$ the experimental results can be accounted for in the framework of the Coffey and Clem model with fluxons moving in the flux-flow regime, provided that the anisotropy of the upper critical field is taken into due account. We have also shown that, at least in the temperature range of about 3 K below $T_c$, the upper-critical-field anisotropy follows the anisotropic Ginzburg-Landau theory. The value of the anisotropy factor of the Alfa-Aesar-powder sample agrees with those reported in the literature for randomly oriented powder of $MgB_2$. However, although the two investigated samples have similar shape, their anisotropy factors are different, corroborating the idea, commonly reported in the literature, that the values of the characteristic parameters of the $MgB_2$ compound depend on the preparation method of the samples. The enhanced field dependence of $R_S$ may be ascribed to both the weak-pinning effects and the relatively low values of the upper critical field of this compound.

## Acknowledgements

The authors are very glad to thank E. H. Brandt, M. R. Trunin, I. Ciccarello, for their continuous interest and helpful suggestions, G. Lapis for technical assistance, N. N. Kolesnikov and M. P. Kulakov for supplying one of the $MgB_2$ samples. Work partially supported by European Fund P. O. N. 940023I1.

**Figure captions**

Fig. 1. Temperature dependence of $(1/Q_L - 1/Q_U)$, for the P#1 and P#2 samples, in the absence of static magnetic field.

Fig. 2 Normalized values of the microwave surface resistance of the P#1 sample as a function of the temperature, at different values of the static field. The normal-state surface resistance, $R_N$, has been determined at $H_0 = 0$ and $T = 40$ K.

Fig. 3 Temperature dependence of the upper critical field deduced from the $R_S/R_N(T, H_0)$ values, for both P#1 (squares) and P#2 (triangles) samples. The inset shows the procedure followed to determine $H_{c2}(T)$.

Fig. 4 Normalized values of the surface resistance as a function of the static magnetic field for P#1 (a) and P#2 (b), at different values of the temperature. Symbols are experimental data. Continuous lines are the best-fit curves, obtained by taking into account the upper-critical-field anisotropy as explained in the text. The lines in (a) have been obtained with $l_0/d_0 = 10^{-2}$, $g = 3$, $H_{c2}^{\perp c}(35) = 10$ kOe, $H_{c2}^{\perp c}(36) = 7.5$ kOe, $H_{c2}^{\perp c}(37) = 3.8$ kOe, $H_{c2}^{\perp c}(38) = 1.7$ kOe. The continuous lines in (b) have been obtained with $l_0/d_0 = 10^{-2}$, $g = 5$, $H_{c2}^{\perp c}(36) = 6$ kOe, $H_{c2}^{\perp c}(37) = 3.3$ kOe. The dashed line has been obtained with $l_0/d_0 = 10^{-2}$ and $H_{c2}(36) = 6$ kOe, disregarding the anisotropy of the upper critical field.

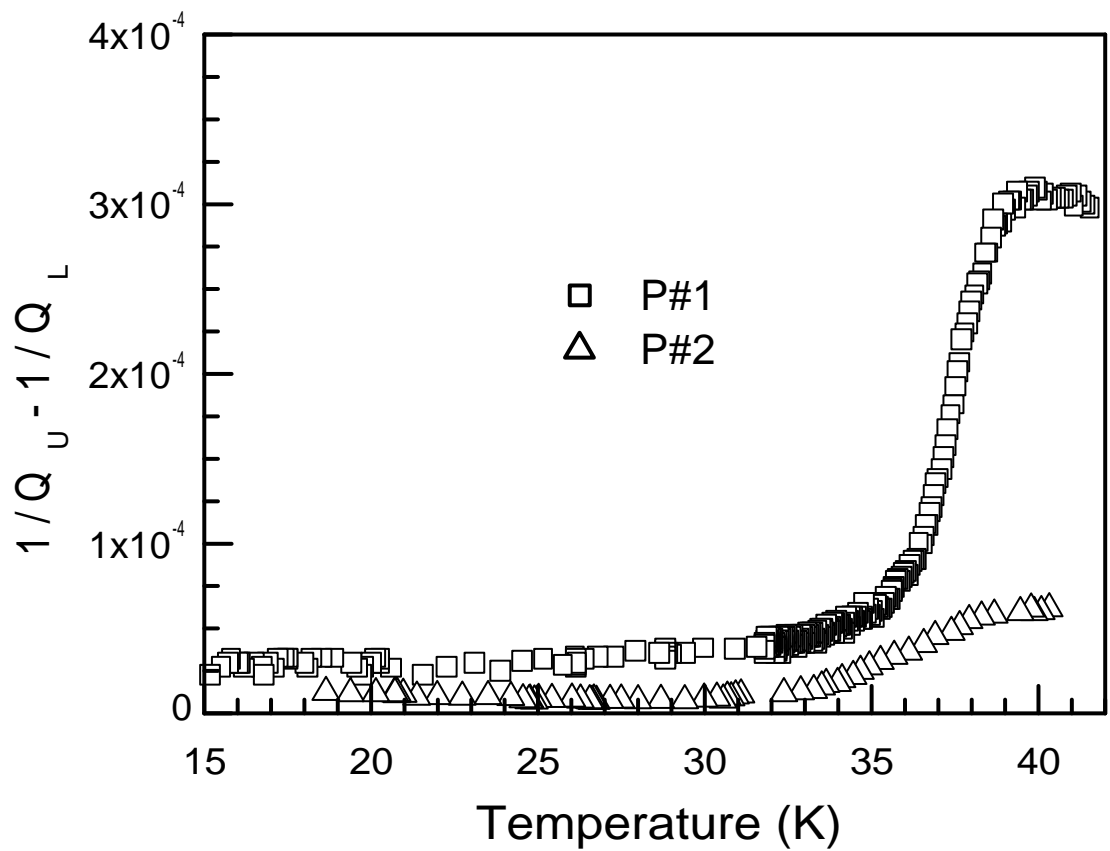

Fig. 1

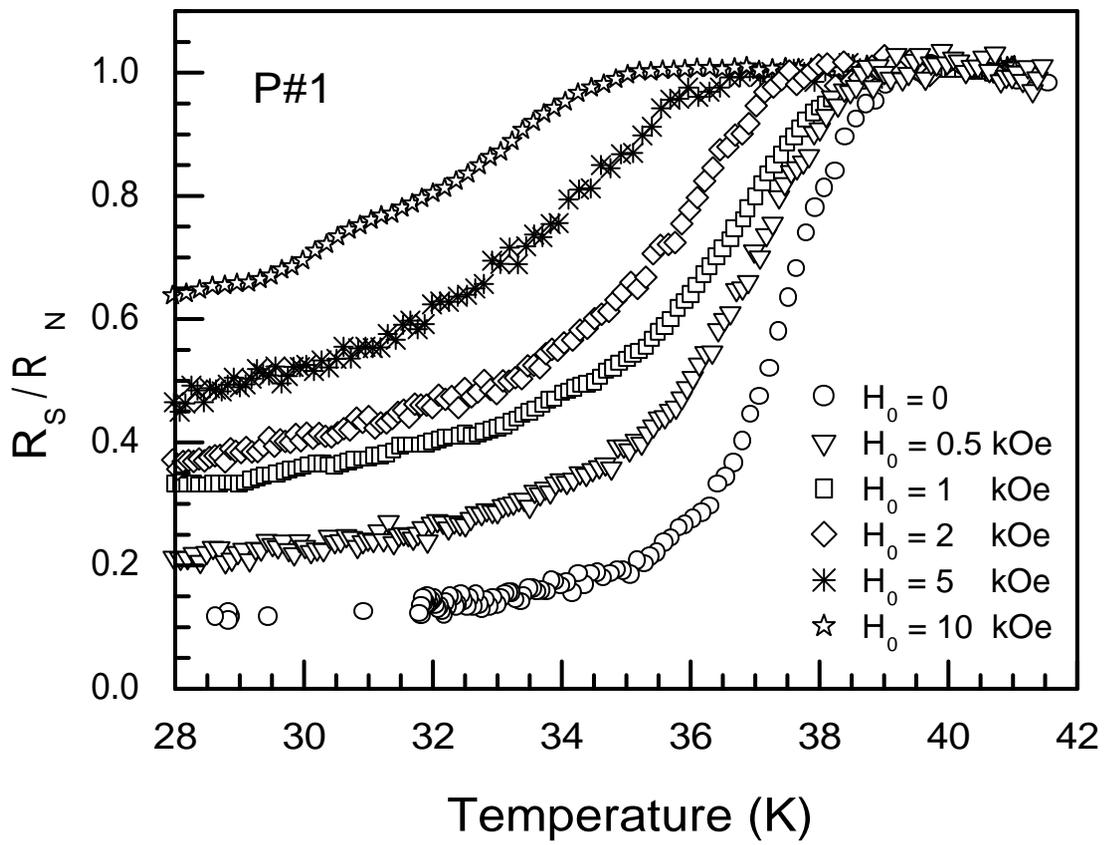

Fig. 2

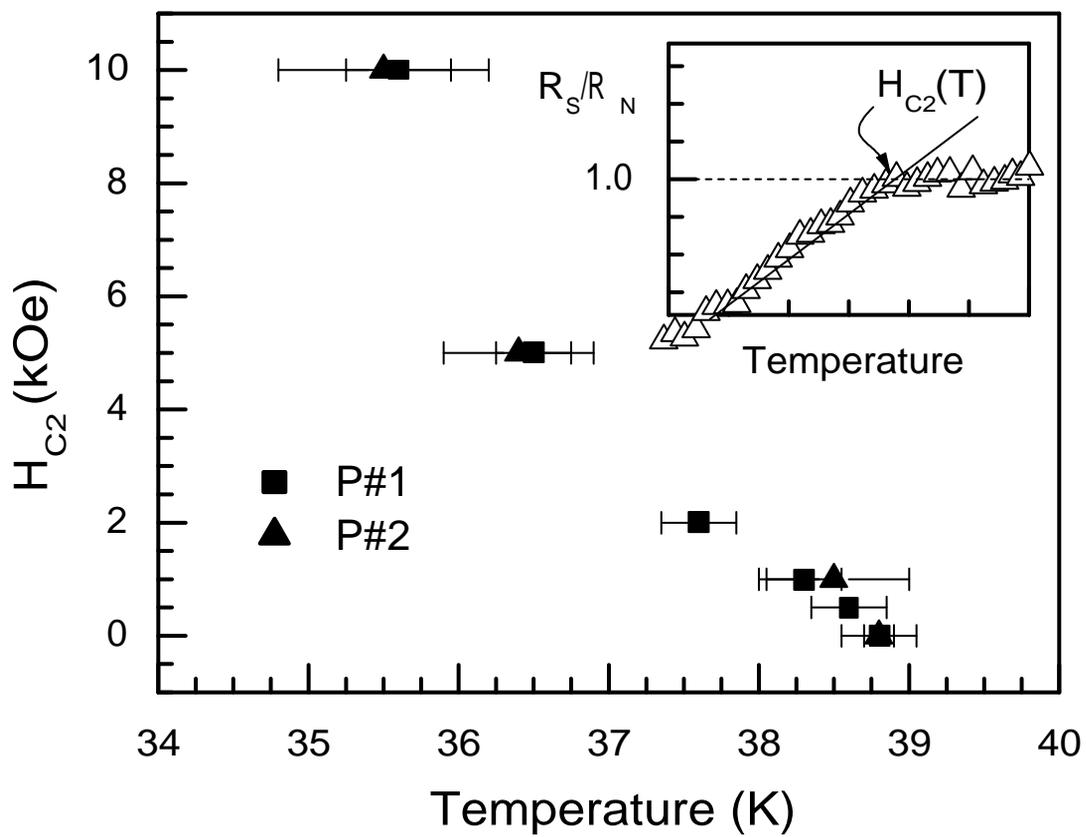

Fig. 3

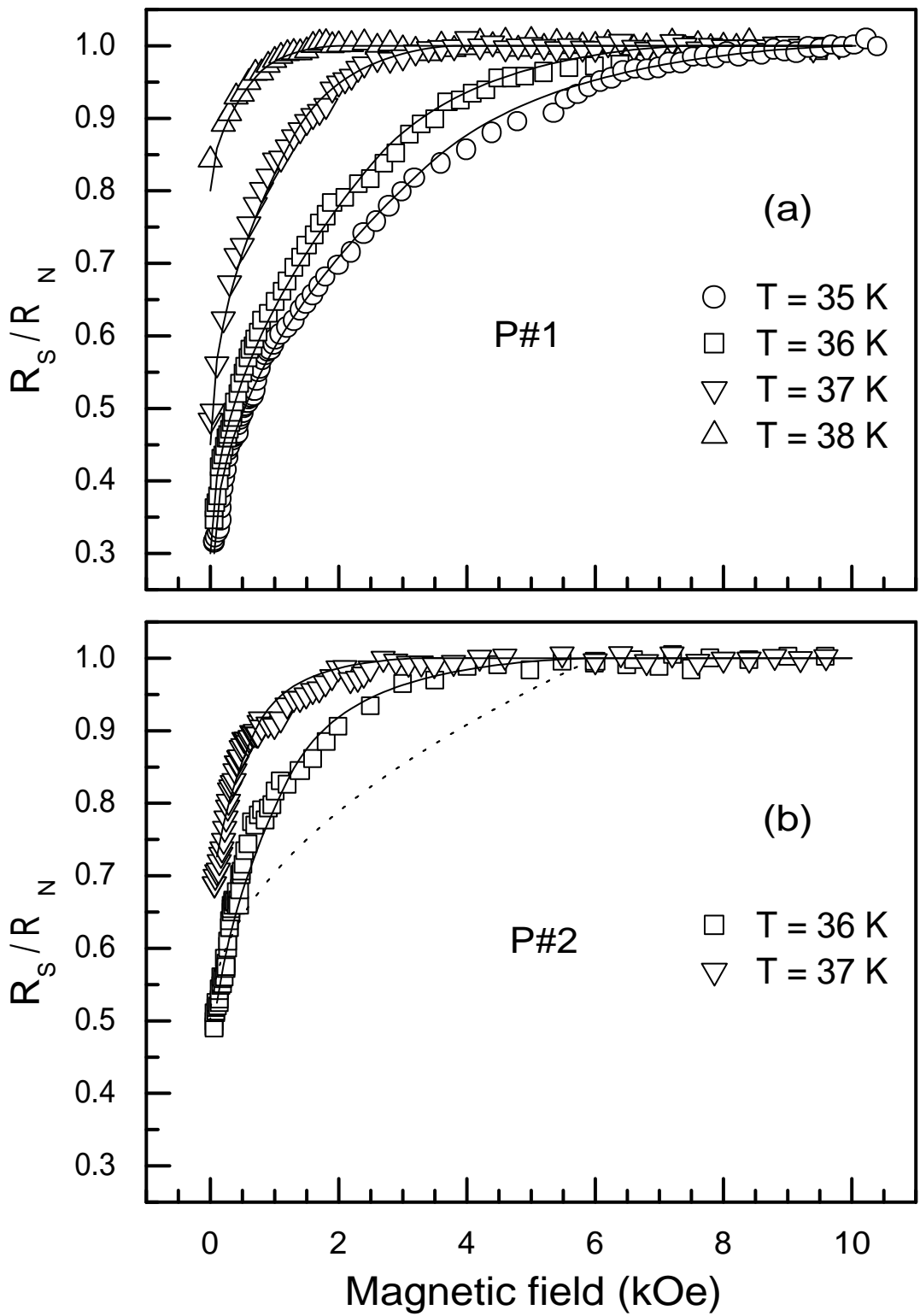

Fig. 4